\documentclass{aa}
\usepackage{graphicx}
\input psfig.tex
\begin{document}

\authorrunning{P. Saracco }
\titlerunning{Deep near-IR observations of the Chandra and HDF-S fields}

   \title{Deep near-IR observations of the Chandra Deep Field and of the 
HDF South\thanks{Based on observations collected 
at the ESO-VLT telescope (Prog. ID.64.O-0643, PI E. Giallongo; 
Prog. ID.164.O-0612, PI M. Franx).}}
	\subtitle{Color and Number Counts}


   \author{P. Saracco$^1$, E. Giallongo$^2$,  S. Cristiani$^{3,5}$, S. D'Odorico$^4$, A. Fontana$^2$,  
 A. Iovino$^1$,  F. Poli$^2$, E. Vanzella$^{4,5}$}

   \offprints{Paolo Saracco, e-mail saracco@merate.mi.astro.it}

   \institute{$^1$Osservatorio Astronomico di Brera, via E. Bianchi 46, Merate,  Italy\\
   $^2$Osservatorio Astronomico di Roma, via Dell'Osservatorio 2, Monteporzio, Italy\\ 
   $^3$Space Telescope European Coordinating Facility,ESO, Karl-Schwarzschildstr. 2, Garching bei Munchen, Germany \\
   $^4$ESO, Karl-Schwarzschildstr. 2, Garching bei Munchen, Germany\\
   $^5$Dipartimento di Astronomia, Universit\'a di Padova,Vicolo dell'Osservatorio, 2, Padova, Italy}
   \date{Received 20 December 2000; Accepted 4 April 2001 }

   \abstract{
We  present near-IR (J and Ks) number counts and colors of galaxies
detected in deep VLT-ISAAC images centered on the 
Chandra Deep Field and Hubble Deep Field-South for a total
area of 13.6 arcmin$^2$.
The  limiting surface brightness obtained is Ks$\simeq$22.8 mag/arcsec$^2$
and J$\simeq$24.5 (1$\sigma$) on both  fields.
A d$log$N/dm relation with a slope of $\sim0.34$ in J and $\sim0.28$
in Ks is found in both  fields with no evidence of decline near the 
magnitude limit.
The median J-Ks color of galaxies  becomes bluer at magnitudes fainter 
than Ks$\sim18$, in agreement with the different number counts slope
observed in the two bands.
We find a fraction ($\le5\%$ of the total sample) of sources 
with color redder than J-Ks=2.3 at magnitudes Ks$>20$.
Most of them appear as isolated sources, possibly elliptical
or dusty starburst galaxies at redshift $z>2$.
The comparison of the  observed  number counts with models shows that
our J-band and Ks-band  counts are consistent with the prediction 
of a  model based on a small amount of merging in a $\Omega=1$ cosmology.
On the other hand, we fail to reproduce the
observed counts if we do not consider merging independently
of the parameters defining the universe.   
      \keywords{cosmology --
                galaxies --
               }
   }
   \maketitle
%


\section{Introduction}
Near infrared (IR) observations provide  significant advantages over optical
ones in studying galaxy evolution because of the small and almost
galaxy type independent k-corrections at these wavelengths and
of their insensitivity to star formation activity over a wide redshift
range (e.g. Cowie et al. 1994).
They also provide a first guess about the stellar mass content 
 of the galaxies since they trace
the underlying old stellar population of galaxies out to $z\sim2-3$.
These are the main reasons why deep near-IR selected samples have been
thought to provide constraints on the different galaxy formation and 
evolution scenarios.

When the first large IR detectors became available, near-IR galaxy
counts were initially considered a powerful cosmological test 
(Cowie et al. 1990; Gardner et al. 1993).
It was later   realized that optical and near-IR
observations were difficult to reconcile without 
number evolution
(Broadhurst et al. 1992; Gronwall \& Koo 1995; Babul \& Rees 1992).

Disentangling the effects of luminosity and number density evolution
is fundamental in understanding the nature of field galaxy formation
and evolution allowing a direct comparison with models.
For instance, hierarchical models, which assume that galaxies assemble
through merging from smaller sub-units, make concrete predictions
about the evolution of the merger rate, morphological mix
and redshift distribution (e.g. Baugh et al. 1996; Kauffmann 1996;
Kauffmann \& Charlot 1998).
In this scenario the number density of massive (and   
luminous in the near-IR) galaxies should decrease with increasing redshift.
Consequently, an IR selected sample is better suited to investigate
the density of these galaxies.

Kauffmann \& Charlot (1998) suggest that the redshift distribution 
of a K-band selected sample can provide an observational test
able to discriminate between hierarchical models and pure luminosity
evolution (PLE) models.
They show that hierarchical models predict a substantially lower fraction
of high-$z$ galaxies at any redshift than do PLE models.
Moreover,  PLE models predict a  high redshift tail of the 
redshift distribution of galaxies much more extended than in hierarchical
models. 
Fontana et al. (1999) applied this test to a composite sample, deriving
the photometric redshift distribution of galaxies down to K$<21$.
They found only $\sim5\%$ of galaxies at $z>2$ in good agreement
with predictions of hierarchical models.

Near-IR colors  have been used,  e.g. by Eisenhardt
et al. (2000), to select $z>1$ galaxies 
with the aim to constrain the fraction of high redshift
galaxies in a K-band selected sample.

We have obtained deep J and Ks band VLT-ISAAC observations 
 centered on the Chandra Deep Field ($\sim6$ arcmin$^2$, CDF hereafter)
and Hubble Deep Field South  ($\sim7.4$ arcmin$^2$, HDFS hereafter)
which  extend to
J$\sim24.5$ and Ks$\sim23$.
The HDFS data are in common with those of a similar proposal.
This data set constitutes the widest area surveyed at these 
near-IR magnitude limits.
In this paper we present counts and near-IR colors of galaxies
detected in these two fields.
The plan of the paper is as follows:
in \S 2  we describe the observations and the data reduction and show 
in \S 3 the procedure adopted to construct our photometric samples
and to derive number counts.
In \S 4 we present color distributions and
in \S 5 we compare the observed number counts  with
different models.
In \S 6 we use color selection to identify
$z>1$ and $z>2$ galaxies and to obtain lower limits to their
number densities.
Finally,  in \S 7 we summarize our results and conclusions.   

\section{Observations and Data Reduction}
\subsection{Observations and Photometric Calibration}
The data have been obtained   with the ISAAC infrared imager/spectrometer
(Moorwood et al. 1999) at the ESO VLT-UT1 telescope.
ISAAC is equipped with a 1024$\times$1024 pixel Rockwell Hawaii array
providing a plate scale of 0.147 arcsec/pix and a total field of view of
 about 2.5$\times$2.5 arcmin.
The two observed fields  are centered  at 03h32m16s, -27$^o$47'25'' 
the center of the Chandra Deep Field and at 22h32m55s, 
-60$^o$33',08''  the center of
the Hubble Deep Field-South.
The observations were gathered over several nights from September to December
1999  under homogeneous seeing
conditions: around 0.6 arcsec in the case of the HDFS and around 0.7 arcsec
in the case of the CDF.
The CDF was imaged in the J and Ks bands while the HDFS in the Js, H and 
Ks bands (see Tab.1).

The standard ``auto-jitter'' mode  was used to take all the images,
with  the telescope being offset by random amounts up to 30 arcsec (CDF)
and 20 arcsec (HDFS) between individual short exposures.
In Tab. 1 we summarize  the number of frames and the
relevant total exposure times in each band for the two observed fields.

Photometric calibration of the observations has been made by observing several
standard stars from the list  of Infrared NICMOS Standard Stars 
(Persson et al. 1998) with magnitudes ranging between 10 and 12.
Instrumental total flux has been estimated by deriving the growth curve
for each star using the IRAF task {\em phot}.
The estimated magnitudes have then been  corrected for atmospheric extinction
assuming A$_{\rm J}$=0.1, A$_{\rm H}$=0.04 and A$_{\rm Ks}$=0.05.  
The typical uncertainty in the derived photometric calibration 
 ranges between 0.02 and 0.04 mag.

\begin{table*}
\begin{center}
\begin{tabular}{lrrrrrr} 
\hline
\hline

Filter & $\bar\lambda$& $\Delta\lambda$& Number of & t$_{\rm exp}$& FWHM   & $\mu$(1$\sigma$) \\
       & ($\mu$n)  &($\mu$n)    &  frames   &   (s)        &(arcsec) & mag/arcsec$^2$   \\
\hline
       &     &       & CDF&      &       &                        \\
\hline
J      &1.25 & 0.26 &67      &12060  &0.65           &24.40 \\
Ks     &2.16 & 0.27 &304     &30400  &0.70           &22.85 \\
\hline
       &        &       & HDFS&           &     &      \\
\hline
Js      &1.24 & 0.16 &210    &25200  &0.59             &24.55 \\
Ks     &2.16 & 0.27 &487    &29220  &0.62           &22.74\\
\hline
\end{tabular} 
\caption{Details of observations and image quality}
\end{center}
\end{table*}

\subsection{Data reduction}
Raw frames have been first corrected for bias and dark current pattern
by subtracting a median dark frame.
The flat field corrections have been made by using a mean differential 
sky-light flat obtained by averaging the difference between a set of 
low-count 
sky flat images  from  a set of high-count sky flat images.
 
Frames have been then processed by DIMSUM\footnote{Deep Infrared Mosaicing Software, 
a package of IRAF scripts by Eisenhardt, Dickinson, Stanford and Ward,
available at ftp://iraf.noao.edu/contrib/dimsumV2.} 
to produce the sky subtracted frames on the basis of the following recipe.
A first pass of DIMSUM produces a ``raw'' final co-added image by which 
to derive a master mask frame flagging  pixels belonging 
to sources.
This master mask is de-registered to create a proper mask relevant to each 
frame.
For each frame a sky background image is thus generated 
by averaging a set (from 3 to 6) of time adjacent frames where sources
are now masked out. 
This procedure makes it possible to obtain sky subtracted frames 
 where sources are not surrounded by  dark halos 
 typical of a sky subtraction process performed without masking 
objects.
 
The sky-subtracted frames have been then inspected and sky residuals
 were removed where present.
Sky residuals have been modeled  using the IRAF task {\em imsurfit}
by fitting a bi-cubic spline to the frame after masking out the sources.

Frames have been then rescaled to the same
airmass using the airmass information stored in the header of the images
and to the same zeropoint by comparing the photometry of all the sources 
detectable in each frame.
Finally, before shifting and co-adding,  frames have been additively rescaled
to the same median value.

Shifting and co-adding have been then performed with DIMSUM.
The final co-added image is the  average of the shifted frames.
The quality of the final images is shown in Table 1 where the measured FWHM
and the 1$\sigma$ limiting surface brightness are reported. 

\begin{figure*}
\centerline{\psfig{figure=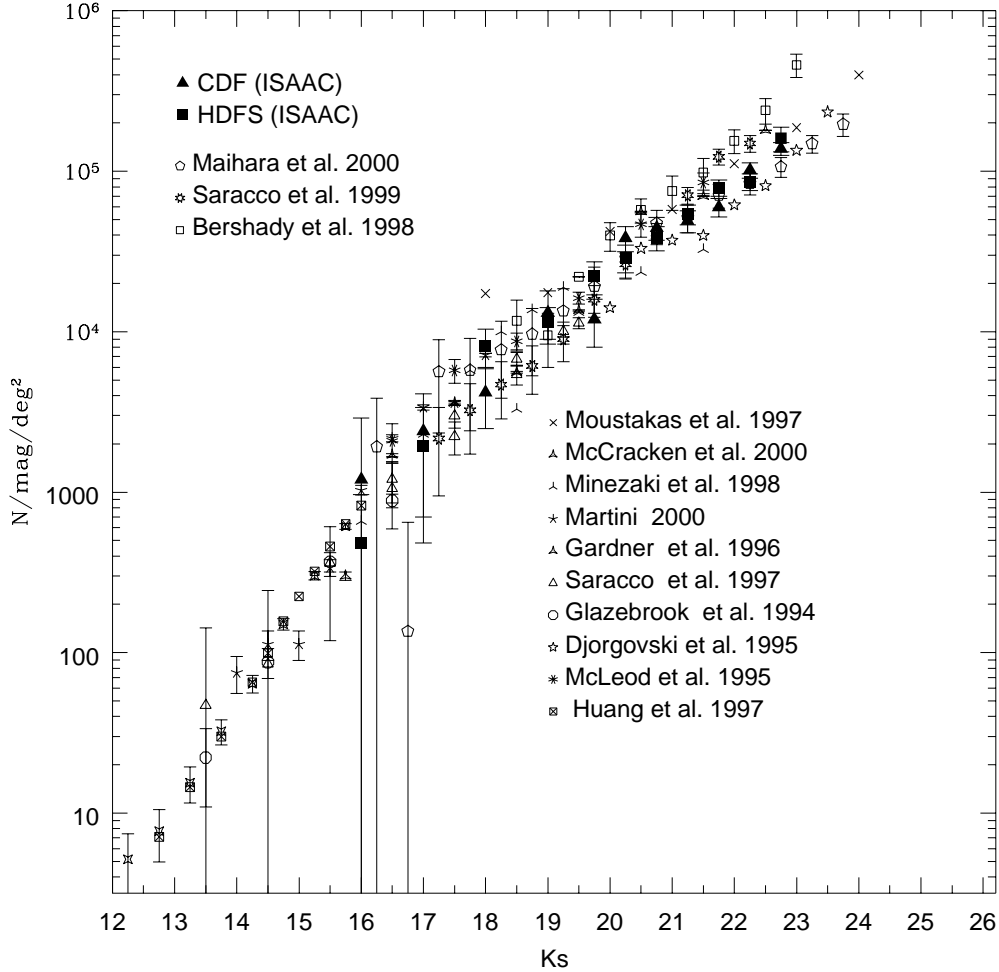,height=140mm}}
\caption{The figure compares the Ks-band counts obtained in this work in 
the CDF and in the HDFS with those in the literature.
The Maihara et al. (2000) counts have been plotted down to their S/N$\sim$3
limit. 
The slope at Ks$>19$ is 0.28 both for 
the CDF and the HDFS Ks-band counts.}
\end{figure*}

\begin{figure*}
\centerline{\psfig{figure=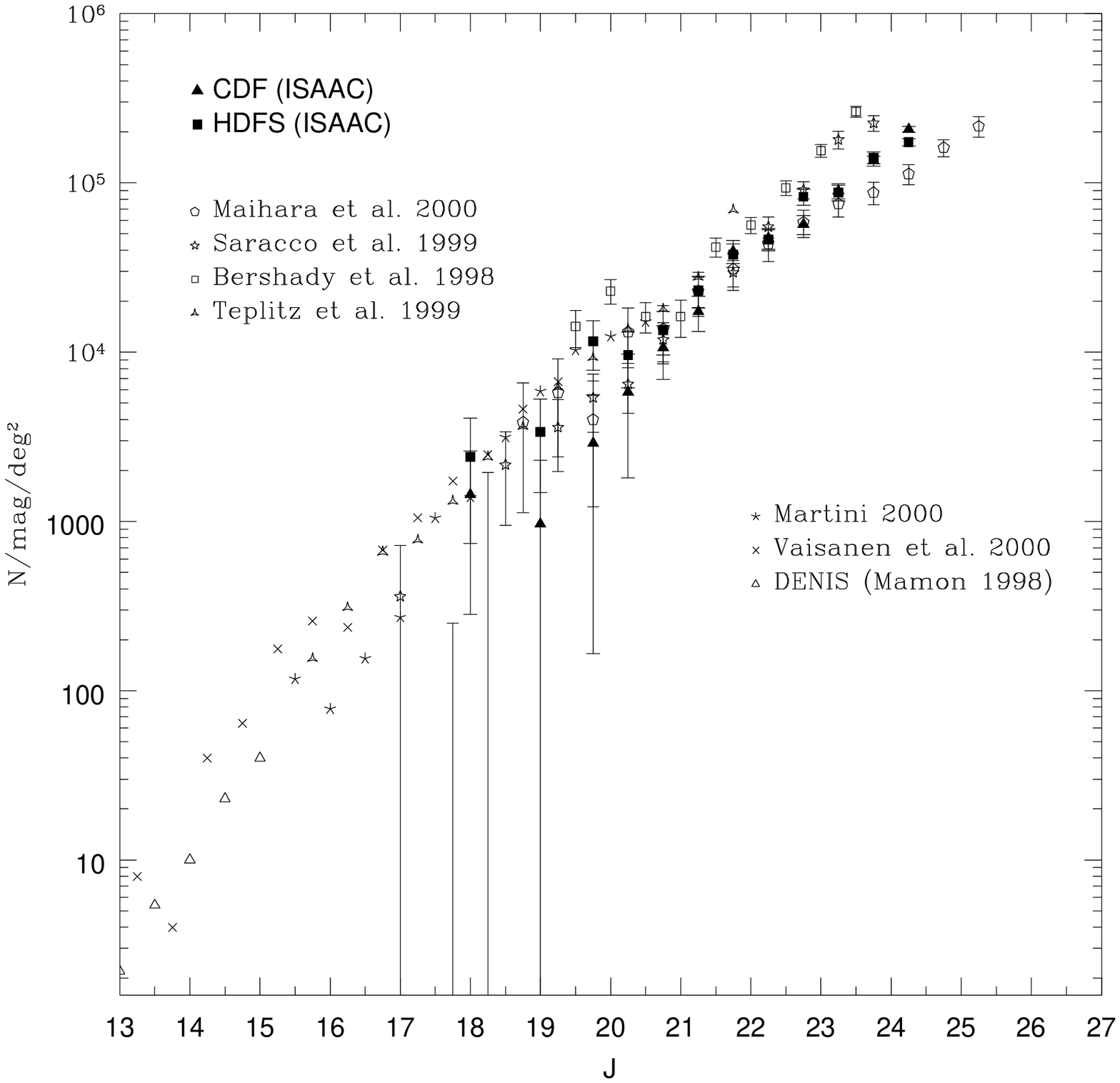,height=140mm}}
\caption{The J-band galaxy counts  obtained in this work are compared with 
those in the literature.
The Maihara et al. (2000) counts have been plotted down to their S/N$\sim$3
limit.  
Our counts continue to rise with a power law slope at J$>20$ of 
0.37 and 0.34 in the CDF and the HDFS respectively.}
\end{figure*}

\section{Number Counts}
\subsection{Object catalogs and magnitudes}

Object detection has been performed using the SExtractor package
(Bertin and Arnouts 1996).
We used a Gaussian function  with a FWHM matching the one measured 
on the frames to convolve the image and an RMS weighting map to optimize
the detection procedure.
Since the two data sets have been obtained under  slightly different seeing
condition, we used  different detection thresholds  such that the faintest 
detectable source in both the fields has a minimum signal-to-noise ratio 
of S/N=3 over the seeing disk. 
The raw catalogs were then been  cleaned of 
those objects having saturated pixels and/or lying too close to the 
edges to perform a reliable flux estimate.

The Ks-band selected  catalogs thus obtained contain 332 sources
over an area of  6.01 arcmin$^2$ (CDF) and 414
sources  over  7.47 arcmin$^2$ (HDFS).
The CDF near-IR catalog is available in electronic form via 
http://www.merate.mi.astro.it/$\sim$saracco.
The multi-band HDFS catalog will be published in a forthcoming paper
(Vanzella et al. 2001).

In order to establish the most reliable and unbiased estimate 
of the "total" flux of sources in our images, we compared different magnitude  
estimates among them.
We focused our attention on three different issues: 1) the 
magnitude which minimizes the systematic underestimate of the flux at
faint magnitudes;
2) identifying the most reliable estimate of the flux of blended sources and 
3)  the best color estimate of the sources.

We made use of the simulated images to compute the completeness corrections
to number counts (see \S 3.2) to systematically investigate
these three different issues.
The results of our analysis   
agree with those previously found by 
Martini (2000) and Saracco et al. (1999) showing
 that the SExtractor BEST magnitude (i.e. isophotal corrected magnitude for 
sources flagged by SExtractor as blended and Kron-like magnitude for the 
remaining sources) underestimates the flux when approaching the limits of the survey.
This bias is much larger in the
case of isophotal magnitudes (both corrected or not).
We also found that a 2 arcsec aperture photometry ($\sim$3$\times$FWHM in our
images) brightened by 0.15 mag (derived by point-like sources)
gives a good estimate of the magnitude of the faintest sources while
underestimate the flux of Ks$<$20.5 sources and fails to recover
that of blended sources.

We thus estimated magnitudes on the basis of the following steps:
\begin{itemize}
\item
the 2 arcsec aperture magnitude corrected by 0.15 mag was assigned
   to those ``isolated'' sources having an isophotal diameter (on the 
   smoothed image) smaller than 2 arcsec.
\item 
the BEST magnitude was assigned to the remaining sources.  
\end{itemize} 
This is very similar to  the  procedure adopted in the previous papers (Fontana et 
al. 2000; Arnouts et al. 1999)

\subsection{Galaxy counts}
To derive galaxy counts, we first removed stars from the samples relying
on the Sextractor star/galaxy classifier.
We defined stars as those sources   having a 
stellarity index larger than 0.9  and a point-source FWHM in all the bands.
A visual inspection made on the F814W band image of the HDFS
confirmed the reliability of the classification.
We found 16 and 23 stars (all of them brighter than Ks=21) in the CDF and in the HDFS 
respectively. 
The possibility that our classification fails to detect fainter stars 
implies negligible differences in the number counts at these
magnitudes, since the surface density of stars is at least one order of
 magnitude lower
than that of galaxies. 

 We have computed the correction factors 
for faint, undetected sources following the recipe described in Saracco 
et al. (1999) and Volonteri et al. (2000).
We generated for each band a set of frames by  dimming our final images 
by various factors and adding pure poissonian noise  at the same 
level as that of the original images.
Sextractor was then run with the same detection parameters to search
for real sources in the dimmed frames.
The correction factor $\bar c$  is the mean number
of dimmed galaxy which should enter the fainter magnitude bin over the mean 
number of detected ones.
In Tab. 2 we report the raw counts $n_r$,  the 
correction factors $\bar c$, the counts per square degree corrected for 
incompleteness and their errors $\sigma_N$.   
Errors were obtained by quadratically summing the Poissonian contribution
$\sigma_{n_r}=\sqrt(n_r)$ of raw counts,  and the uncertainty on the 
correction factor $\sigma_c$.
\begin{table*}
\begin{center}
\begin{tabular}{lrrrrrrrr} 
\hline
\hline
   &        &            & CDF &6.01 arcmin$^2$&                  &           & HDFS    & 7.47 arcmin$^2$       
               \\
\hline
Ks & $n_r$  &  $\bar c$ & N/mag/deg$^2$ & $\sigma_N$ & $n_r$ &  $\bar c$ & 
N/mag/deg$^2$ & $\sigma_N$\\
\hline
16.00&  3    &  1 &   1198  &	    1194 &   1 &     1  &   482   &	   482\\
17.00&  4    &  1 &   2396  &	    1694 &   9 &     1  &  1928   &	  1446\\
18.00&  12   &  1 &   4193  &	    1694 &  21 &     1  &  8193   &	  2208\\
19.00&  33   &  1 &   13180  &       4792 &  29 &     1  &  11566   &	  2595\\
19.75&  11   &  1 &   11980  &       3973 &  25 &     1  &  22168  &	  5161\\
20.25&  32   &  1 &   38340  &       6777 &  30 &     1  &  28916  &	  5653\\
20.75&  37   &  1 &   44330  &       7287 &  40 &     1  &  38554  &	  6528\\
21.25&  41   &  1 &   49120  &       7671 &  56 &     1  &  53976  &	  7724\\
21.75&  44   &  1 &   52710  &       7947 &  82 &     1  &  79036  &	  9346\\
22.25&  77   &1.1 &   101500 &  1.103e+04 &  79 &  1.13  &  86043  &	 10370\\
22.75&  35   &3.3 &   138400 &  1.288e+04 &  37 &   4.5  & 160482  &	 28250\\
\hline
\hline

J & $n_r$  &  $\bar c$ & N/mag/deg$^2$ & $\sigma_N$ & $n_r$ &  $\bar c$ & 
N/mag/deg$^2$ & $\sigma_N$\\
\hline

 18.00  &    3  &   1	&   1445 &  1163 &  6	  &    1  &   2409&   1669\\
 19.00  &   5	&   1	&    963  &  1339 &  10    &	1  &   3373&   1894\\
 19.75  &   8	&   1	&   2891 &  2726 &  15    &    1  &  11566&   3732\\
 20.25  &   17  &   1	&   5783 &  3974 &  13    &    1  &   9638&   3475\\
 20.75  &   15  &   1	&  10602 &  3732 &  16    &    1  &  13493&   3855\\
 21.25  &   18  &   1	&  17349 &  4089 &  26    &    1  &  23132&   4914\\
 21.75  &   41  &   1	&  39518 &  6171 &  39    &    1  &  37590&   6019\\
 22.25  &   49  &   1	&  47228 &  6746 &  48    &    1  &  46265&   6677\\
 22.75  &   59  &   1	&  56867 &  7403 &  87    &    1  &  82891&   8990\\
 23.25  &   93  &   1	&  89638 &  9295 &  91    &    1  &  87710&   9194\\
 23.75  &   86  &   1.3 & 134891 &  8938&  133    &   1.1 & 141092&  12300\\
 24.25  &   70  &   2.5 & 207469 &  8064 &  86    &  2.1 &  174000&  18938\\
\hline
\hline

\end{tabular} 
\caption{Differential number counts in the Ks-band and in the J-band
derived on the CDF (upper panel) and on the HDFS (lower panel).
Errors take into account the Poissonian error on raw
counts and the uncertainties in the completeness correction factor
$c$.}
\end{center}
\end{table*}

Fig. 1 and 2 show the number counts derived for the Ks-band  and
for the J-band, respectively.
We do not find significant deviations between the CDF and HDFS.
The counts follow a d$log$N/dm relation with a slope $\gamma_J\simeq0.34$ at 
J$\ge20$ and $\gamma_K\simeq0.28$ at Ks$\ge19$.
Also plotted in the figures are  number counts derived from the previous 
surveys taken from the literature.

The surface density of galaxies we derived  as well as the
rate at which it increases with apparent magnitude, i.e. 
the slope of the counts, lie in between those derived previously by other 
surveys. 

The remarkable scatter among the surveys may be ascribed to the 
superposition of various effects: the slightly different filters (K, K', Ks
and J, Js), the different method used to estimate magnitudes, possible
systematics in the photometric calibration and  cosmic variance
which could significantly affect the number counts at faint magnitudes where 
the area covered by the surveys is always narrower than few arcmin$^2$.
It is actually at magnitudes fainter than K=21 and J=23 that the largest
discrepancies among various surveys occur.
Finally, fields centered on high-redshift target objects may
result in a significant excess of the number counts (e.g. Soifer et al. 1994).

The CDF and HDFS Ks-band counts derived here are systematically lower 
than those of Bershady et al. (1998), Saracco et al. (1999) and McCracken 
et al. (2000).
 The largest discrepancy occurs  with respect to Bershady's 
counts (a factor  1.7 at Ks$\sim22.5$) which  increase 
faster (especially in the Ks band) with respect to our counts,  
the slope being  $\sim0.36$.
{ Both Bershady et al. (1998) and McCracken et al. (2000) surveyed  
high Galactic latitude blank fields  covering an area smaller than 2 arcmin$^2$.
They count more than 200 and 80 objects down to K$\sim23.5$ and 
K$\sim22.5$ respectively.
While Bershady et al.  use fixed aperture magnitudes (within $\sim2$ arcsec) corrected
to total on the basis of the object size,  McCracken et al. use a Kron-type magnitude. 
Saracco et al. (1999) surveyed an area of  20 arcmin$^2$ 
centered on the NTT Deep Field (Arnouts et al. 1999)  and including
the  high-redshift quasar BR 1202-07. 
They detect $\sim1000$ objects to a limiting magnitude K$_s\sim 22.5$ and apply
an aperture correction to the 2.5 arcsec fixed aperture magnitude.
The method we used to estimate magnitudes  
is quite similar to the one  used both by Bershady et al. and Saracco et al.
The different surface density of galaxies they  found with respect to the one here 
derived may thus be ascribed to cosmic variance affecting the small area covered
by Bershady et al. and the field  surveyed by 
Saracco et al. due to the presence of the high-redshift quasar. 
In the case of McCracken et al. data  their different method used
to estimate magnitudes adds to the high count fluctuations expected 
in such a small area.
}

On the other hand our counts, which agree with those of Moustakas et al. 
(1997), are systematically higher than those of Djorgovski et al. (1995)
and of the recent very deep counts derived by Maihara et al. (2000)
on the Subaru Deep Field (SDF).
In particular the CDF and HDFS counts are in excess at magnitudes J$\ge23$
and Ks$\ge22$ with respect to the SDF counts which 
are described by a slope of $\sim0.23$ { and they are always $\sim1.4$ times 
higher than those derived by Djorgovski et al.
}
 Bershady et al. (1998) claim that the corrected aperture magnitude
 used by Djorgovski et al. (1995) may result in an overestimation of the depth
of their survey up to 0.5 mag.

It is worth noting that extrapolating the  scaling relation
found for field K-band selected galaxies (Roche et al. 1999)
to Ks$\sim$21, we found that clustering can produce count fluctuations
of $\sim25\%$ in the magnitude range $20.5<Ks<21.5$
 over an area comparable to  the CDF and HDFS.
These fluctuations rise up to $\sim30-32\%$ over  areas of 2-1 arcmin$^2$.
Thus,  large scale structure fluctuations alone could in principle 
account for most of the observed discrepancies  among the surveys at 
faint magnitudes.

\begin{figure}
\centerline{\psfig{figure=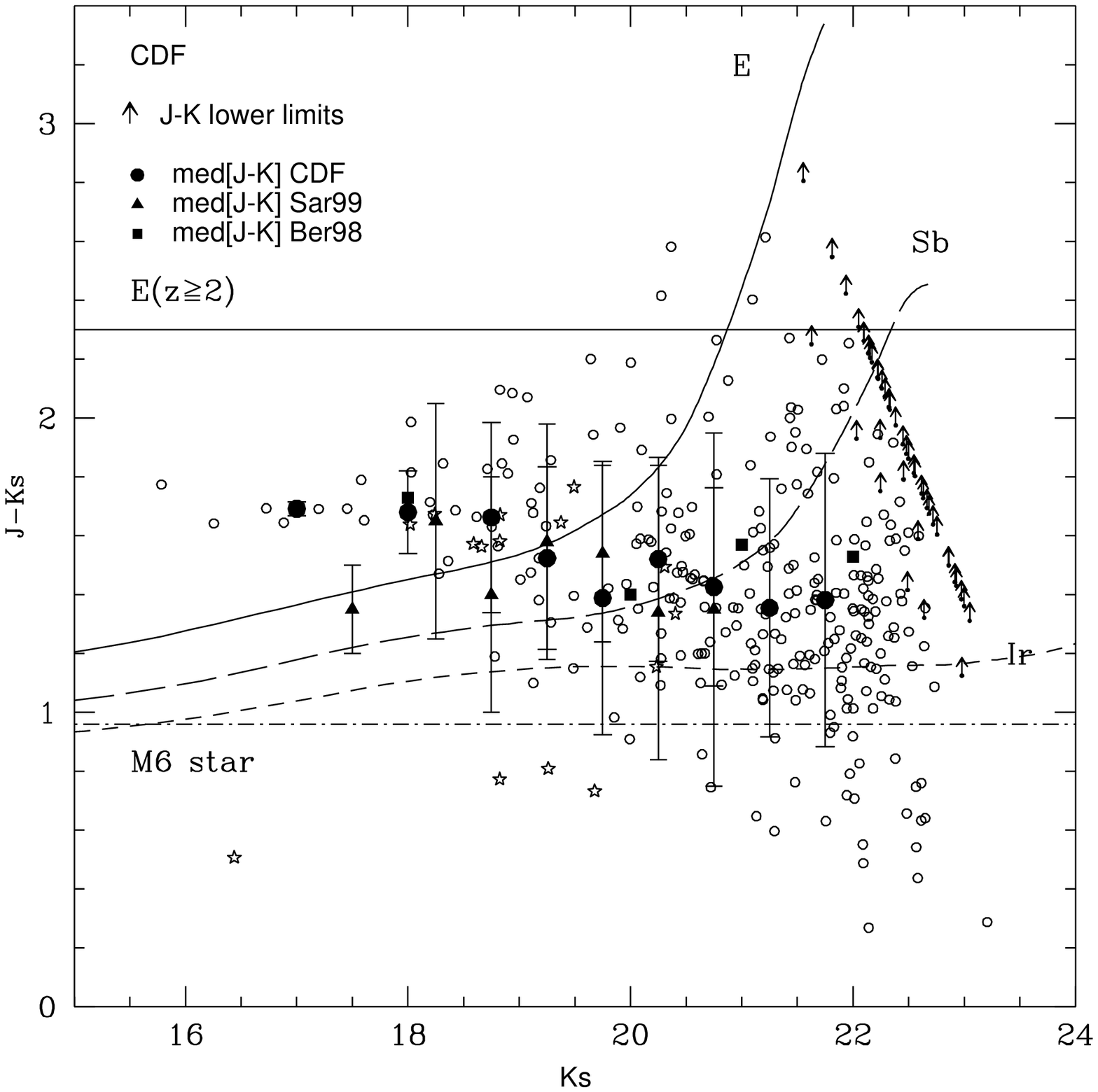,height=90mm}}
\caption{Color-magnitude diagram for the CDF Ks selected sample. Open circles
represent galaxies detected both in the J and in the Ks band  the arrows
mark the J-Ks lower limits.  The horizontal lines represent the typical J-K color of
a main sequence M6 star (dashed line, J-K$\simeq0.96$) and of an elliptical
galaxy at $z\ge2$ (solid line) respectively. The expected J-Ks color for
a pure galaxy population of ellipticals (E, solid curve), spirals 
(Sb, long-dashed curve) and irregular (Irr, short-dashed curve) types is 
also shown (H$_0$, $q_0$ = 50, 0.5).}
\end{figure}
\begin{figure}
\centerline{\psfig{figure=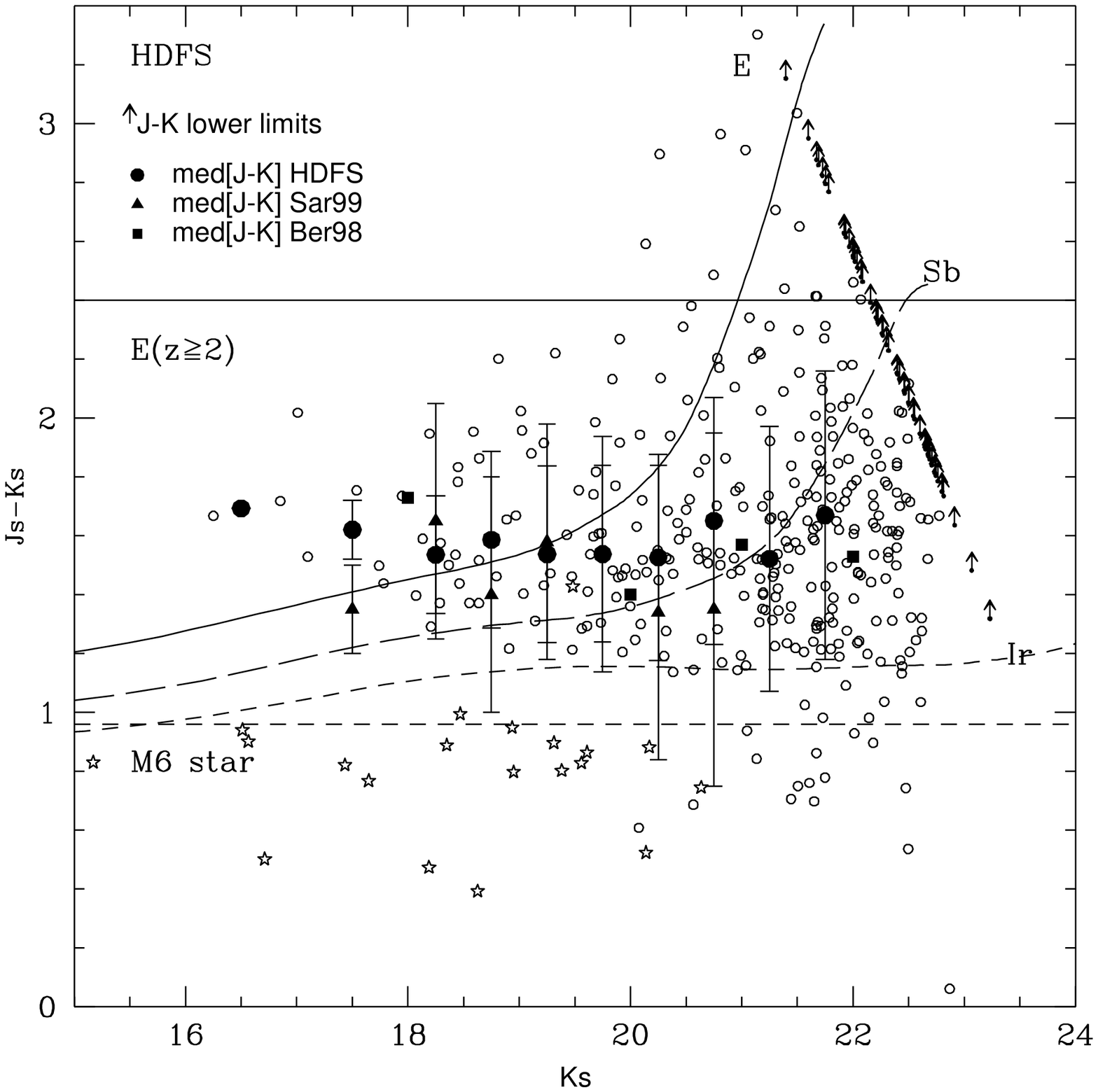,height=90mm}}
\caption{Same as Fig. 3 but for the HDFS sample}
\end{figure}

\section{Colors of galaxies}
The near-IR color of galaxies of our samples has been obtained by running 
SExtractor in the so-called {\em double-image mode}: J (Js) magnitudes for 
the Ks
selected sample have been derived using the Ks frame as a reference image
for detection and the J (Js) image for measurements.
We used the difference between the 2  arcsec diameter aperture magnitudes 
($\sim3\times$FWHM) to estimate the color of isolated sources while the 
difference
between isophotal magnitudes has been adopted to measure the color of
blended sources.
This is to minimize the uncertainties on flux estimates occurring for 
blended sources in the case of a fixed  aperture magnitude.
We  considered genuine estimates those signals exceeding 2$\sigma$ above
the background while fainter signals have been replaced by the 2$\sigma$ 
upper limit.

In Fig. 3 and 4 the color-magnitude diagrams of the Ks selected CDF and HDFS
samples are shown respectively: open circles represent galaxies detected both 
in the J and in the Ks band while the lower limits to the J-Ks color are 
represented by the arrows.
{ It is worth noting that the different depth reached in the J 
filters (see Tab. 1) gives rise to a  difference of $\sim0.15$ mag 
in the limiting J-Ks color between the two fields.} 
Point-like sources are marked by star symbols.
The median J-Ks color in 0.5 magnitude bins for
our sample (filled circle) is also plotted and compared to the value given
in  Bershady et al. (1998, filled squares) and Saracco et al. 
(1999, filled triangles).
Error bars on the median values are the standard deviation within each bin.
The horizontal lines represent the typical J-K color of
a main sequence M6 star (dashed line, J-K$\simeq0.96$) and of an elliptical
galaxy at $z\ge2$ (solid line).
In the figures we also plotted the expected J-Ks color of an $M^*$ 
elliptical (solid curve), spiral (Sb, long-dashed curve) and 
irregular (short-dashed curve) galaxy in the redshift range $0<z<3$. 
 Model prediction are based on Buzzoni's
(1989, 1995) population synthesis code: ellipticals are
described by a single burst stellar population, spirals are modeled
taking into account a declining star formation rate in the disk
coupled with a spheroidal  component as in the elliptical and irregulars
follow a flat star formation rate at every age.

\begin{figure}
\centerline{\psfig{figure=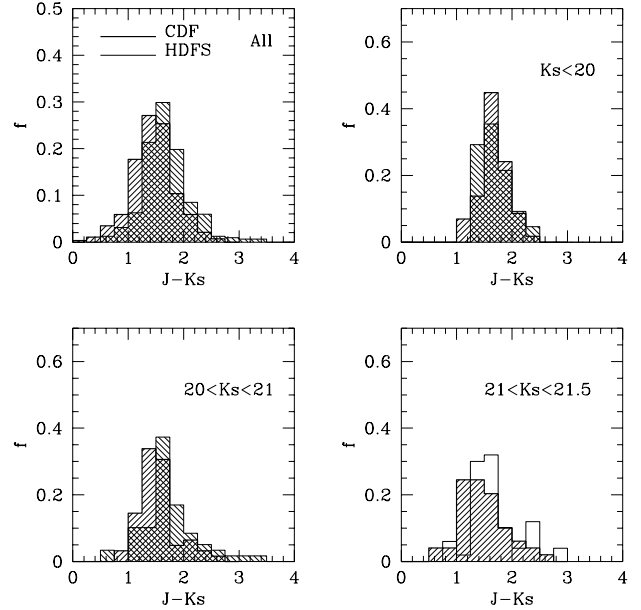,height=90mm}}
\caption{J-Ks color distributions for the  whole CDF and HDFS sample 
(All) and for subsamples selected in different Ks magnitude ranges. 
At bright magnitudes (Ks$<20$) the two samples are described by the same 
color distribution.
At fainter magnitudes the  HDFS  
sample shows colors  slightly redder  than  to those of the CDF one.}
\end{figure}

We have compared the color distributions of the two samples:
a feature present in the CDF sample is the bluing trend described
by the median  color occurring at magnitude Ks$\ge18$ where the J-Ks color
change from $\sim1.7$ at Ks$\sim18$ to $\sim1.4$ at  Ks$\sim21.5$.
The median colors agree well with those in Bershady et al. (1998) and 
Saracco et al. (1999) as well as the bluing trend we found, a trend  much 
more evident in the case of optical-IR colors (see e.g. Gardner et al. 1993; 
McCracken et al. 2000).
 In the HDFS sample the bluing trend is less obvious.
At bright magnitudes (Ks$\le20$) the CDF and HDFS samples follow
almost the same color distribution while at fainter magnitudes the median 
colors in the HDFS seem to remain constant and   are 
slightly but systematically redder than those in the CDF.
This is shown in Table 3 where we report the median colors for the two
samples as a function of Ks magnitude and in Fig. 5 where the color 
distributions in different Ks magnitude slices for the CDF and HDFS samples
are compared.
{ The different behavior shown by  the 
median color of the two samples is not due to the different
limiting J-Ks color between the two fields, at least down to Ks=21.5
Indeed, in this magnitude range all the K-band selected objects of 
both samples have
been detected in J.
}

{ The median color in the CDF   seems to be slightly
redder than in the HDFS at very bright magnitudes (Ks$<19$)
while it is 0.1 bluer than the HDFS at Ks$>19$.
These small differences cannot be 
ascribed to systematics in  photometric calibration since most 
of the observations were been carried out during the same nights.
Various effects can in principle contribute to this difference.
For instance, the presence
of an intermediate-$z$ cluster in such  small areas would strongly 
influence the median colors.
 This could be the case of the CDF where the presence of
a cluster at $z\sim06-0.7$ (Cimatti, 
private communication) would affect colors at magnitudes Ks$<19$, giving rise to
the slightly redder color observed in this interval.
At Ks$>19$ the larger number of red (J-Ks$>1.9$) objects
populating  the HDFS with respect to  the CDF (see Fig. 3 and 4) 
gives rise to the 
different median colors.
This different number of red objects can be explained in terms of ERO
 clustering 
(see \S6).}
Finally, the use of slightly different filters  
could produce systematics in the color distributions.   
 The standard J filter used to image the CDF is  slightly 
redder and much wider than the Js filter (see Tab. 1) and   has some leaks 
in the K band.
All these features act to make the CDF sources bluer than the HDFS,
an effect more evident going to high redshift where the rest-frame optical part of
the spectra enters the J filter.
{ We have tried to quantify this effect by convolving  the  response 
functions of the two filters
with a synthetic spectrum of an Sb galaxy devoid of emission lines.
At $z=0$ the difference between the estimated magnitudes within the two 
filters is 0.012 mag in the sense that the galaxy is brighter in J
(bluer in J-Ks), as we expected.
At redshift $z=2$, where the Balmer break enters the J filter, the difference
rises  to 0.083 mag.
The possibility that an emission line falls in the J filter
but not in the Js due to its narrower width significantly increases 
this difference.
For instance, the presence of a typical H$\alpha$ emission line 
(which enters the J band at $z\sim0.9$) would add
0.04 mag to the magnitude in the J filter.

\begin{table}
\begin{center}
\begin{tabular}{lrr|rrr} 

\hline
\hline
    Ks	&Js-Ks&$\sigma$&    Ks &J-Ks&$\sigma$\\
        & HDFS  &      & & CDF   & \\
\hline
   16.50&  1.74& 0.01&    --&	    --&  --\\
   17.50&  1.67& 0.1&  17.00&      1.74& 0.02\\
   18.25&  1.58& 0.2&  18.00&	   1.73& 0.1\\
   18.75&  1.64& 0.3&  18.75&	   1.71& 0.3\\
   19.25&  1.59& 0.3&  19.25&	   1.57& 0.3\\
   19.75&  1.59& 0.4&  19.75&	   1.44& 0.4\\
   20.25&  1.58& 0.3&  20.25&	   1.57& 0.3\\
   20.75&  1.70& 0.4&  20.75&	   1.48& 0.3\\
   21.25&  1.57& 0.4&  21.25&	   1.40& 0.4\\
   21.75&  1.72& 0.4&  21.75&	   1.43& 0.5\\
\hline		       
\hline
\end{tabular}
\caption{Median J-Ks color of galaxies as a function of magnitude.
Errors are the standard deviation in each bin.}
\end{center}
\end{table}

\section{Comparison with models}
We have compared our number counts  with different models 
by using $ncmod$, the
publicly available code  by Gardner (1998).
The aim of this comparison is not to define how galaxies
evolve, since it is well known that number counts alone are not
able to put severe constraints on the evolution of galaxies.
We would rather be interested in understanding 
the most probable mechanisms able to reconcile optical and near-IR
number counts in the light of the latest
estimates of $\Omega$.

We have considered two families of H$_0=70$ Km s$^{-1}$ Mpc$^{-1}$ 
flat cosmological models:
the first one defined by 

\noindent
($\Omega_M,\Omega_\Lambda$)=(1.0,0.0) and the second 
 by ($\Omega_M,\Omega_\Lambda$)=(0.3,0.7), according to the recent 
results of the Boomerang and MAXIMA experiments (De Bernardis et al. 2000; Balbi et al. 2000) and in agreement with the results
on Type Ia supernovae (Riess et al. 1998; Perlmutter et al. 1999).
The expected Ks and J galaxy counts for the two scenarios are shown 
in Fig. 6 and 7 respectively.
All the models  are based on the following assumptions:
1)the local K-band luminosity function (LF) and mix of galaxies  of Gardner et al. (1997);
2)  the galaxy spectral energy distributions from GISSEL96 
(Bruzual and Charlot 1993)
We considered five morphological types of galaxies: 
E/S0, Sa, Sbc, Scd and Irr. 
We used in the computation the filter transmission curves of the VLT-ISAAC
J and Ks filters.
In the models, all galaxies formed at $z=15$ with the exception of irregulars
which are always 1 Gyr old.
E/S0, Sab and Scd  evolve following  an exponential 
star formation rate (SFR) with $e$-folding time $\tau=1.0$ Gyr, 
$\tau=4.0$ Gyr and $\tau=7.0$ Gyr respectively while Scd and Irr are assumed 
to be described  by a constant star formation.
These are the basic parameters describing the pure luminosity evolution
(PLE) ``ev'' models
represented in the figures by the short-dashed curve.
These models tend to underestimate the counts at faint magnitudes, 
an effect  much heavier  in the ($\Omega_M,\Omega_\Lambda$)=(1.0,0.0)
case.  
\begin{figure}
\centerline{\psfig{figure=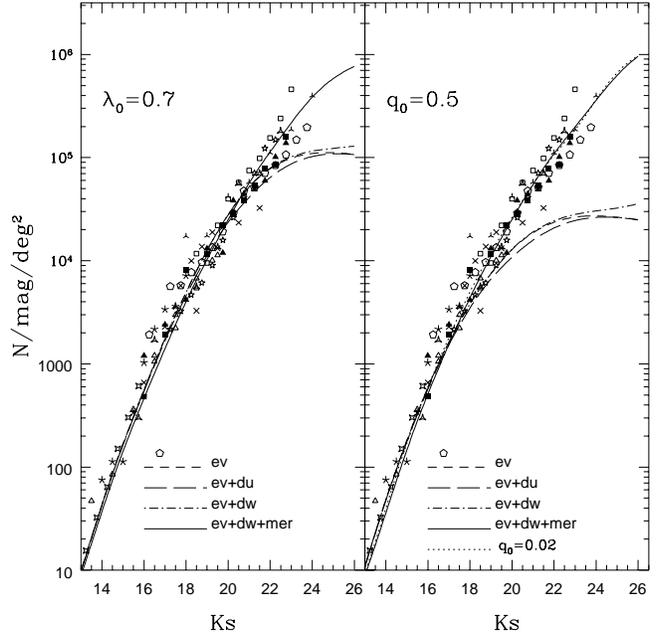,height=90mm}}
\caption{Galaxy count models ($H_0=70$ Km s$^{-1}$ Mpc$^{-1}$) are 
compared to the observed Ks-band counts. 
The long-dashed and short-dashed lines represent a 
PLE model with (ev+du) and without (ev) absorption by dust. 
The addition to the ev model of a dwarf population ($\alpha=-1.6$ for Irr)
is described by the dot-dashed line (ev+dw) while the solid-line represents
a model including merging (ev+dw+me). 
The left-hand panel shows these  models in a 
($\Omega_M,\Omega_\Lambda$)=(0.3,0.7) universe, while the right-hand 
panel considers a ($\Omega_M,\Omega_\Lambda$)=(1.0,0.0) cosmology.}
\end{figure}
\begin{figure}
\centerline{\psfig{figure=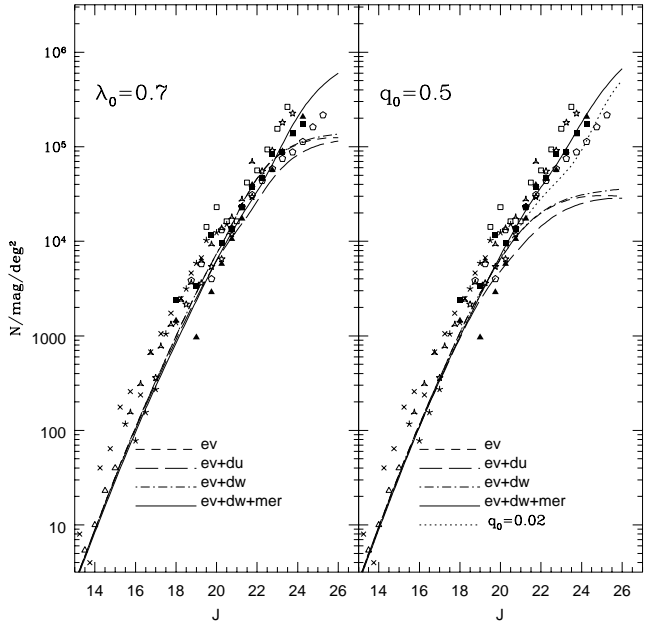,height=90mm}}
\caption{Same as Fig. 6 but for the J-band galaxy counts.}
\end{figure}
{ We will now explore the various standard possibilities 
to reduce the observed excess of
 faint galaxies in the near-IR with respect to our PLE model.}

The addition of dust extinction (long-dashed curve) has been introduced 
in modeling galaxy counts  mainly to reduce the UV excess in galaxies
and to match the U and B-band galaxy counts. 
In this case it enhances the observed excess at faint magnitudes
even if the effect is minimal in the near-IR. 
In these models the absorption by dust internal to the galaxies
is based on the recipe of Wang (1991) and Bruzual et al. (1988)
who modeled the dust as a layer with variable thickness
symmetric around the center of the galaxy.  

A steep faint end LF ($\alpha=-1.6$) describing  the latest types of 
galaxies reduces in principal the discrepancy at faint magnitudes.
Evidence of a steep LF of field galaxies at infrared wavelength
come from Szokoly et al. (1998) who find a slope $\alpha=-1.3$
and from Bershady et al. (1999) who estimate $\alpha=-1.6$.
Other evidence in favor of $\alpha\ll-1$ come from the results
  obtained on optically selected samples of field
galaxies (Marzke et al. 1994, 1997; 
Zucca et al. 1997) and of cluster galaxies (De Propris et al. 1998;
 Lobo et al. 1997; Bernstein et al. 1995; Molinari et al. 1998). 
However, such a steep LF, which contributes substantially to
the galaxy counts at optical wavelengths (see e.g. Volonteri et al. 2000),
affects only marginally galaxy counts
in the near-IR due to the blue optical-IR colors of
irregular and dwarf galaxies. 
The small contribution to the counts given by this steep LF at faint 
IR magnitudes is shown  in the figures by the dot-dashed curve (ev+dw),
a model which assumes  $\alpha=-1.6$ for the LF of irregular galaxies.

We finally considered a model including  both merging and the dwarf component
(solid curve).
This model follows  the number evolution proposed by Broadhurst et al. (1992)
who parameterized merging in the form
$\phi^*\propto$ exp[$-Q/\beta((1+z)^{-\beta}-1)$], where Q 
defines the merger rate and $\beta=1+(2q_0)^{0.6}/2$ (Peacock 1987).
This formalism is preferable to that where merging proceeds
exponentially with $(1+z)$ (e.g. Rocca-Volmerange \& Guiderdoni 1990) 
because the rate increases smoothly at high redshift and does not become 
unreasonably high  at early times.
  
This model fits very well the observed J and Ks galaxy counts 
both in the $\Omega_M,\Omega_\Lambda$=1.0,0.0 and 
$\Omega_M,\Omega_\Lambda$=0.3,0.7 cosmology, assuming a merger rate Q=1.
{ However, the same fit could be  obtained without
the dwarf component by enhancing  the rate of merging to Q=1.5.}

It is worth noting that all the models without merging here considered
predict a turnoff 
in the counts (at Ks$\sim20$ in the q$_0=0.5$ models and at Ks$\sim21$ 
in the  $\lambda_0=0.7$ models), contrary to the observations.
On the other hand, models including merging both in the $q_0=0.5$ and
  $\lambda_0=0.7$ cosmology provide a good fit to the data over the whole
magnitude and wavelength ranges considered. 
The $q_0=0.5$ model including merging and  dwarf component is in fact 
the same model previously explored by Volonteri et al. (2000)
which provided a  good fit also to the HDFS galaxy counts at optical
wavelengths.

{ The merging rate required to fit  our data (from Q=1 to Q=1.5) 
implies that $\Phi^*(z=1)_{Q=1}=1.5\Phi^*(z=0)$ and 
$\Phi^*(z=1)_{Q=1.5}=2\Phi^*(z=0)$,
i.e. a galaxy should undergo  on average  0.5-1 merger events 
from $z=1$ to $z=0$.
 These rate of merging are comparable with those 
derived from 
the observed merger fraction based on counts of
close pairs of galaxies and from the merger identification studies.
The results of Roche et al. (1999) suggest that a galaxy will undergo 
from 0.4 to 0.7 merger events in the redshift interval 
$0<z<1$ while  Le Fevre et al. (2000) 
estimate from 0.8 to 1.8 merger events in the same redshift range.  
Comparing the rate of merging derived by galaxy count models with 
that predicted by a
hierarchical clustering scheme is less obvious. 
The model of Baugh et al. (1996) predicts that 50$\%$ of the elliptical 
galaxies and 15$\%$ of the spiral galaxies have had a major merger in the 
redshift interval $0<z<0.5$.
These numbers increase to $\sim90\%$ and $\sim50\%$ at $z<1$. 
On the other hand, a significant fraction of present day 
ellipticals ($\sim40\%$) should be the result of the merging of four or more
fragments at $z\sim0.5$, while most of the spirals should contain only 
 two  such fragments at $z\sim1$.
The simple galaxy counts  model here considered accounts for a mean merging 
rate averaged over the whole population of galaxies, which at $z\sim1$
should be $\sim1.5-2$ times larger than that at $z=0$.
These numbers seem to be lower than those predicted by hierarchical
clustering.}

In the right-hand panels of figures 6 and 7 is also shown for
interest a PLE $q_0=0.02$ model (dotted-line) including the absorption
by dust internal to the galaxies.
It fits reasonably well both J and Ks-band galaxy counts. 
This kind of model, which seems to provide a good fit to the data over
a large wavelength range (e.g. Pozzetti et al. 1996, 1998; 
Metcalfe et al. 2000; McCracken et al. 2000), is in fact ruled out by 
the recent measurement of $\Omega$.  

\section{Color selections}
\subsection{The fraction of $z>1$ galaxies} 
Kauffmann and Charlot (1998) showed that hierarchical galaxy formation
makes substantially different predictions about the   redshift distribution 
of a K-band selected sample  with respect to  PLE models.
In particular, hierarchical models predict a much smaller fraction of
galaxies at $z>1$ where most of the massive galaxies should not be
already assembled.
The  fraction of $z>1$ galaxies in a K-band selected 
sample can thus represent a powerful test to discriminate
between the different galaxy formation models.

Fontana et al. (1999, 2000) use a photometric redshift technique to derive
the redshift distribution of a  K$<21$ selected sample
of 319 galaxies.
They find a good agreement with the prediction of hierarchical models
and estimate a fraction of 35-40$\%$  of galaxies at $z>1$ and of 5$\%$
at $z>2$.

Eisenhardt et al. (2000) suggest that the J-K color of galaxies is 
a good photometric redshift indicator, showing that all but one of
the galaxies redder than J-K=1.9 in  their spectroscopic sample of
$\sim50$ galaxies are  at redshift $z>1$.
Such a color threshold should actually  select old stellar systems 
at redshift $z>1$,
J-K$\sim1.9$ being  the color of a present day elliptical galaxy at $z>1$.
Consequently, this color selection  provides a lower limit to the fraction 
of $z>1$ galaxies, since later-type $z>1$ galaxies would be discarded
due to their bluer color.
Eisenhardt et al. find a lower limit of 25$\%$ for the fraction of $z>1$ 
galaxies in the EES K$=20$ limited sample.
This number is in agreement with the photometric redshift distribution
of Fontana et al. (2000) at the same limiting magnitude.
Indeed, a careful examination of the published photometric
redshift catalogs used by Fontana et al. (2000) shows that the two
criteria are roughly equivalent at K$<20$, as expected since at these
bright K limits most of the selected galaxies are early type.
  
As a first attempt to extend this test at fainter magnitudes, 
we have applied the selection criterion suggested by Eisenhardt et al. (2000)
to our Ks selected samples in order to derive the lower limit to the
fraction of $z>1$ galaxies.
In the HDFS sample we applied this color selection taking
into account the different J filter used, i.e. 
restricting the selection to those sources redder than Js-Ks$>2$
 (see \S 4).

\begin{table}
\begin{center}
\begin{tabular}{l|rrr|rrr} 

\hline
\hline
    Ks	&N$_{tot}$&N$_{z>1}$&f$_{z>1}$& N$_{tot}$&N$_{z>1}$&f$_{z>1}$\\
        & & HDFS  &      & & CDF&     \\
\hline
   16-17& 5   &0  &$>$0$\%$ &4   &0  &$>$0$\%$\\
   17-18& 10  &1  &$>$10$\%$&4	&0   &$>$0$\%$\\
   18-19& 29  &1  &$>$3$\%$ &26  &4  &$>$15$\%$\\
   19-20& 41  &4  &$>$10$\%$&28  &5  &$>$18$\%$\\
   20-21& 70  &14 &$>$20$\%$&61  &8  &$>$13$\%$\\
   21-22& 138 &37 &$>$27$\%$&99  &19 &$>$19$\%$\\
\hline
\hline
\end{tabular}
\caption{Lower limits to the fraction of $z>1$ galaxies as a function of Ks magnitude.}
\end{center}
\end{table}
In Table 4 the total number of galaxies (N$_{tot}$),
the number of galaxies redder than J-Ks=1.9 (N$_{z>1}$) and their 
fraction with respect to the total (f$_{z>1}$) are shown for each 
Ks magnitude  bin.
We estimate a fraction of J-Ks$>$1.9, Ks$<20$ objects of
14$\%$ and 7$\%$ in the CDF and HDFS respectively, corresponding to  
a surface density of 1.5 and 0.8 objects per square arcmin.
 Eisenhardt et al. find a mean surface density of
J-Ks$>$1.9 objects of 3.4 over the 124 arcmin$^2$ of the EES survey.
They show  that this value can range from 1 to 6.7 on
subfields having a size comparable to the HDF. 
The values we find are thus consistent with those derived by Eisenhardt et al.
 and the differences between CDF and HDFS confirm their findings.

At fainter magnitudes ($20< Ks < 22$), the fraction of J-Ks$>1.9$ objects
rises by 13-27$\% $ in the two fields. 
These numbers are comparable to the
prediction of hierarchical models originally computed by Kauffmann and
Charlot (1998), and lower than the fractions predicted by the revised
hierarchical models computed by Fontana et al (1999), that range from
23$\%$ of the Standard CDM to 40$\%$ of the $\Lambda$ CDM (both at K$<21$).
 At the same time, they are significantly lower than the fraction of $z>1$
predicted by PLE models, which predict up to 75$\%$ of galaxies in the
same magnitude range (Kauffmann and Charlot 1998). Taking into account
the expected fluctuation due to the small size of the field, and
considering that the J-K$>1.9$ criteria produces a lower limit to the
actual fraction of $z>1$ galaxies, we conclude that both the hierarchical  and
PLE models are still compatible with our findings, and that more
accurate determinations of the redshift distributions at $z>1$ are
required at these faint magnitudes to discriminate between them.

\subsection{ EROs selection at $z>2$}
In \S 4 we derived the J-Ks color distribution of the Ks selected CDF 
and HDFS samples noting the presence of some sources redder than J-Ks$\sim2.3$
at magnitudes Ks$>20$.
This is the k-corrected color of an elliptical galaxy at $z\ge1.9$ or
at $z\ge2.1$, in the case of passive evolution (Buzzoni 1995).
This color threshold would thus in principle select old stellar
systems at $z>2$ where the Balmer jump would be redshifted
red-wards of the J filter and the UV emission would be depressed
by the lack of young stellar population. 
On the other hand, similar  colors  also could be displayed
by dusty star-forming galaxies which would look so red
due to the heavy  absorption of the UV emission by dust.

\begin{figure}
\psfig{figure=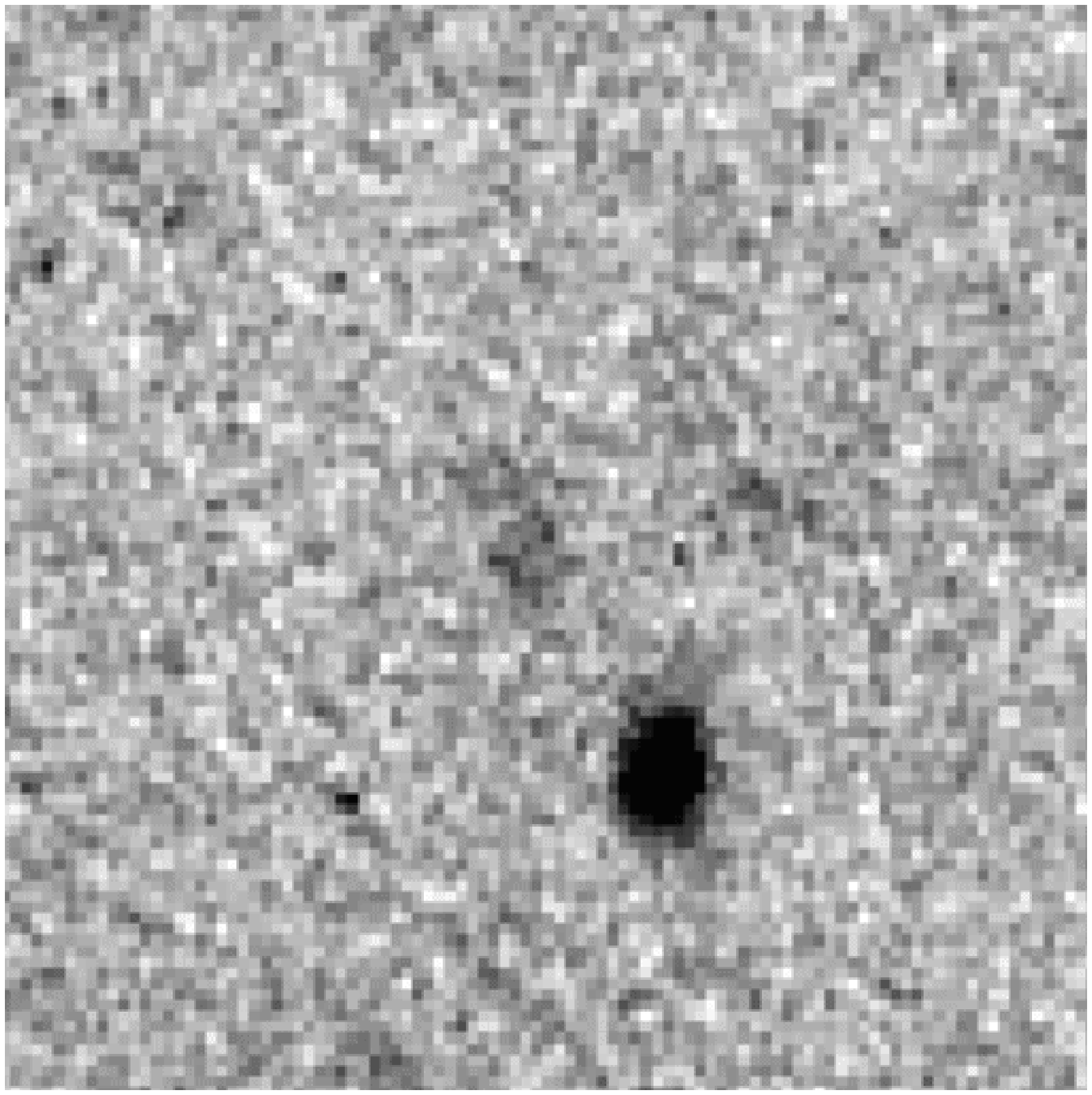,height=35mm}
\vskip -3.5truecm
\hskip 4.5truecm \psfig{figure=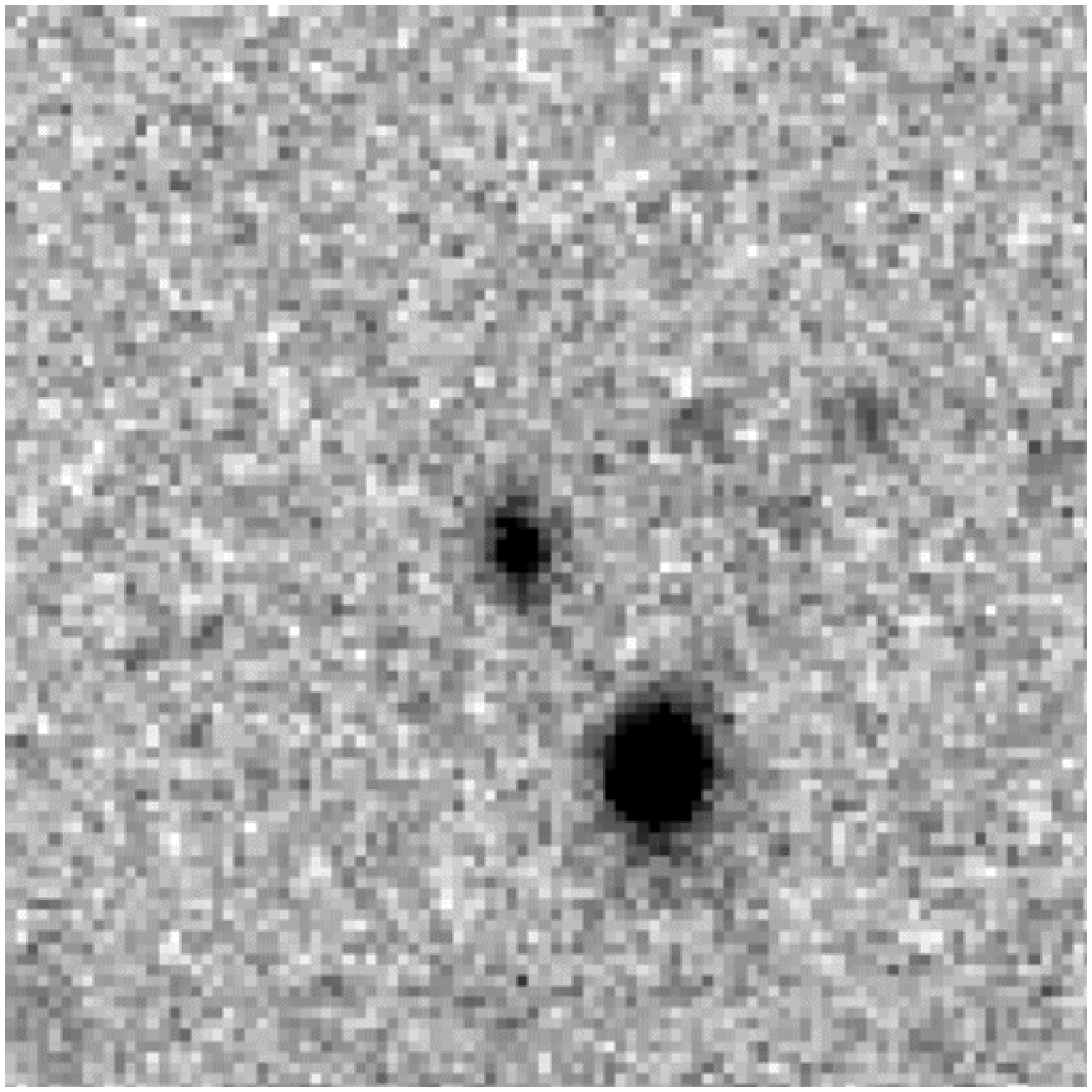,height=35mm}
\caption{The figure shows the J-band image (left-hand panel) and
the Ks-band image (right-hand panel) of source CDF-293.
Both the images are 15$\times$15 arcsec centered on the source
(RA=03:32:13, Dec=-27:46:41)
This source is characterized by
a color J-Ks=2.6 and an apparent magnitude Ks=20.4.}
\end{figure}

\begin{figure}
\psfig{figure=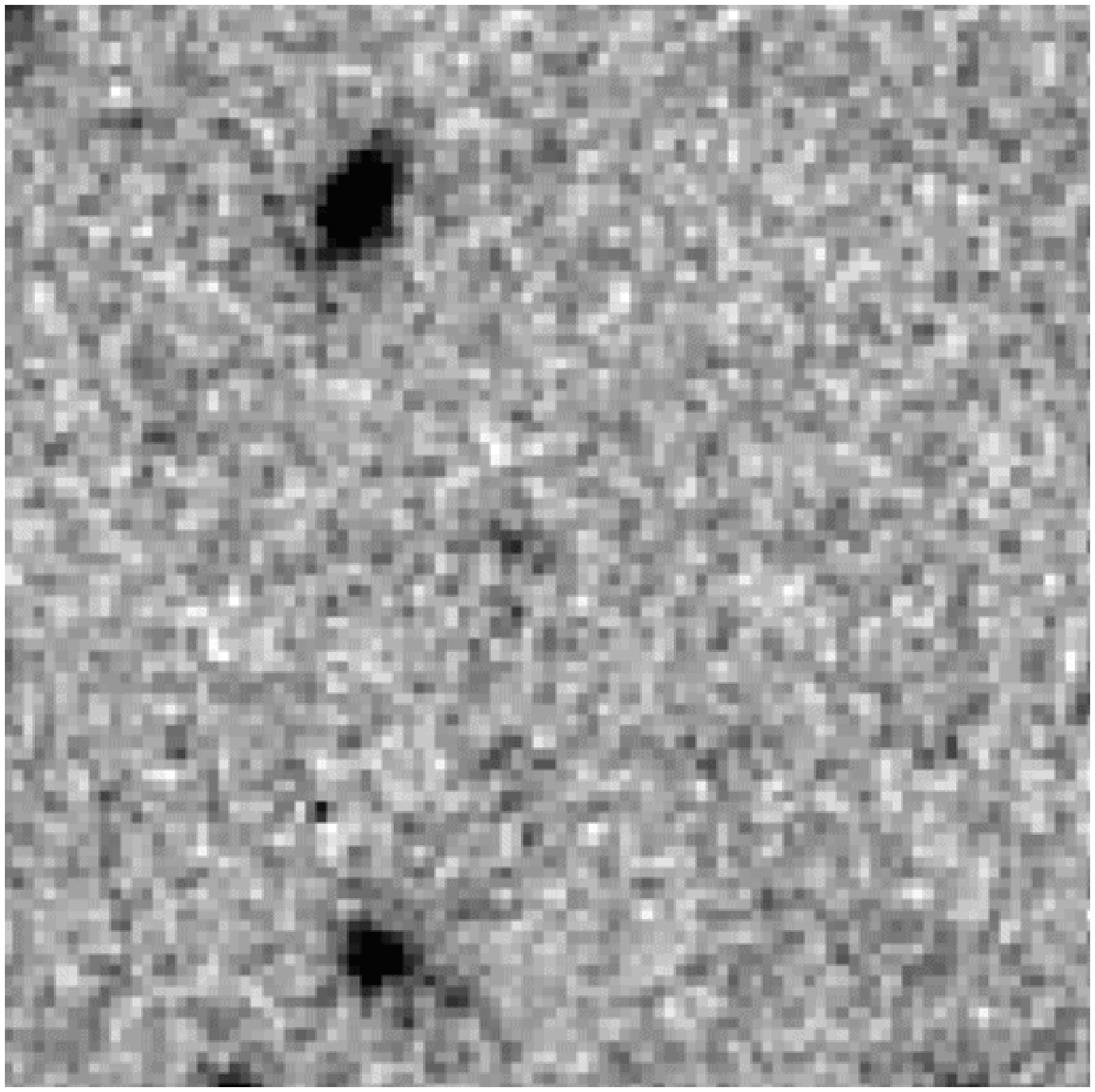,height=35mm}
\vskip -3.5truecm
\hskip 4.5truecm \psfig{figure=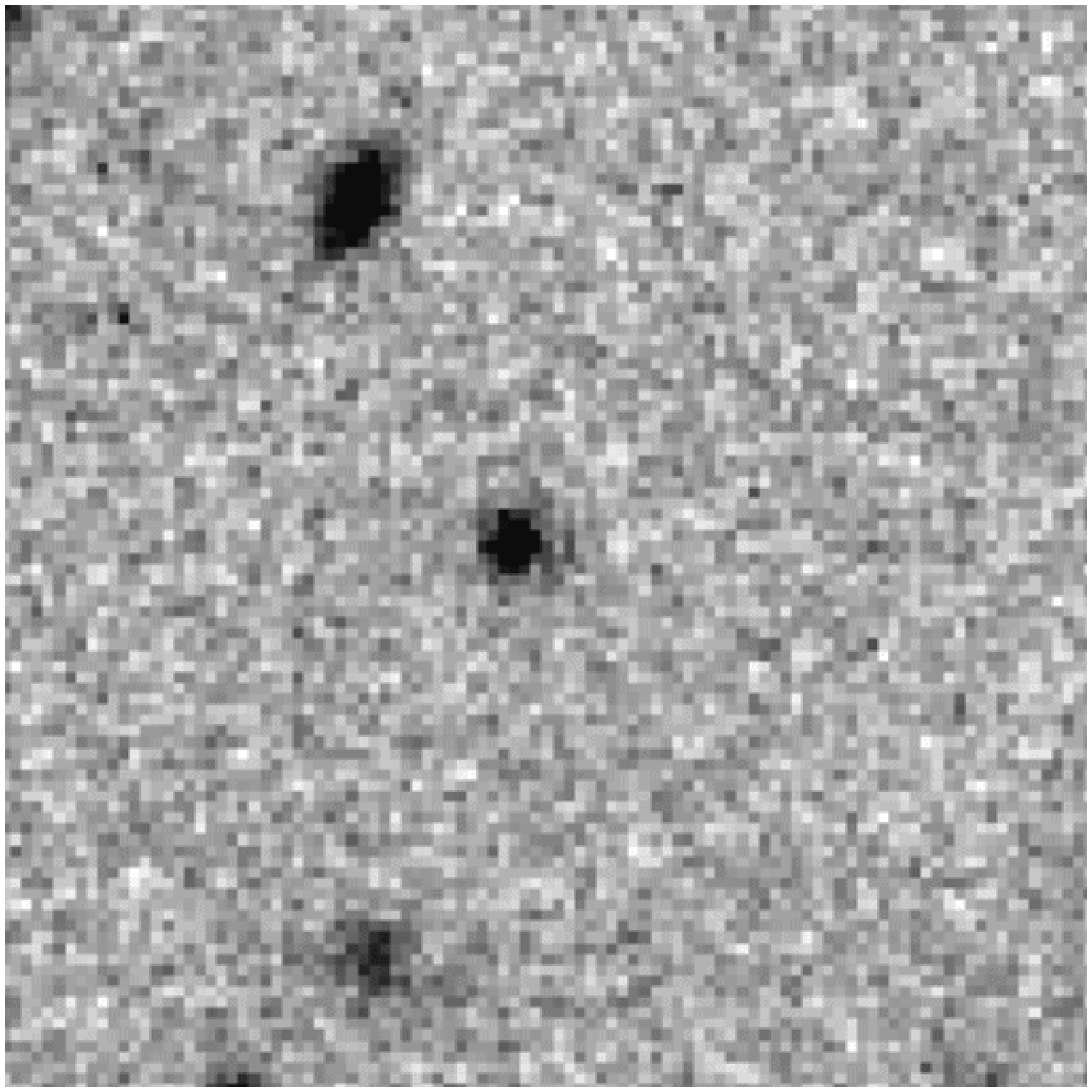,height=35mm}
\caption{The J-band image (left-hand panel) and
the Ks-band image (right-hand panel) of source HDFS-363
are shown.
Both the images are 15$\times$15 arcsec centered on the source
(RA=22:32:49, Dec=-60:32:11).
This source is characterized by
a color Js-Ks=3.3 and an apparent magnitude Ks=20.8.}
\end{figure}

In the CDF sample we count 7 sources  (2$\%$ of the sample)
redder than J-Ks=2.3 at Ks$\le22$,
i.e. a surface density of about 1.2 per square arcmin.
In the HDFS sample we  selected 20 sources (5$\%$ of the sample)
 redder than Js-Ks=2.4 at Ks$\le22$,
equivalent to a surface density of 2.7 per square arcmin.
Six out of these 20 EROs lie out of the WFPC2 field and four on the borders.
Four out of  the remaining ten have no optical counterpart in the F814W
band image.

The different surface densities found are most probably due
to the clustering properties of EROs which introduce a strong field-to-field 
variation in small areas.
Indeed, Daddi et al. (2000) detect a strong clustering signal 
of EROs selected on the basis of the R-K color from a
K$\sim19$ limited sample (see also McCarthy et al. 2000).
They show that such clustering produces fluctuations 
on the surface densities of K$<19$ EROs 1.7 times higher than those expected
from pure poissonian statistic on areas of the order of
 $\sim25$ arcmin$^2$.

We selected EROs on a sample 3 magnitudes fainter and, in principle,
on a different redshift range since we have used the J-Ks color selection.
Thus, the results of Daddi et al. (2000) cannot be directly applied
to our data.
However, we could assume that the ERO clustering
amplitude scales with K magnitude accordingly to the amplitude 
found for field K-band selected galaxies.
On the basis of the results obtained by Roche et al. (1999)  we thus expect 
that at Ks$\sim21$  the ERO clustering is 1.25 times weaker than that 
derived by Daddi et al. at Ks$\sim$19.
In the presence of a correlation with amplitude A, the fluctuations $\sigma$
of the counts around the mean value $\bar n$ are 
$\sigma=\sqrt{\bar n(1+\bar nAC)}$, being $C\propto Area^{-0.4}$.
Applying this relation to our data, we derive an ERO
surface density fluctuation    $\sigma\sim0.7$ 
which could be consistent with the fluctuations we observe.

In Fig. 8 and 9 the J and Ks-band images of the EROs CDF-293
and HDFS-363 are shown as an example. 
These sources have a J-Ks color  of 2.6 and 3.3 respectively
and an apparent Ks magnitude of 20.4 and 20.8.
If they are  elliptical galaxies at redshift  $z\ge2$, their k-corrected 
absolute magnitude would be $M_{Ks}<-25$,
consistent with a passively evolved $M^*(z=0)$ galaxy. 
   
Objects with extreme J-K color fainter than K$>20$ have
been also found in the HDF-N  by Dickinson et al. (2000) and
in the HDF-S NICMOS field by Yahata et al. (2000). 
They discuss the possibility that they are dusty star-forming
galaxies or ellipticals at $z>2$ or $z>10$ Lyman break galaxies.
Maihara et al. (2000) notice that most of the
red sources they found in the SDF appear as close neighbors,
thus being probable interacting systems.
We notice that $\sim30\%$ of the red sources we found both in the
CDF and in the HDFS are flagged as ``blended'' by SExtractor
and that the remaining $70\%$ do not appear as close neighbors.
A final possibility remains that some of the most compact sources
at Ks$>21$ are faint very low-mass stars, which can display 
colors redder than J-K=2 (Chabrier et al. 2000; Dahn et al. 2000).

The analysis of the near-IR properties combined with those derived
by photometry at optical wavelength   will be presented in a forthcoming
 paper (Saracco et al. 2001).

\subsection{Extremely blue sources}
{ Figures 3 and 4 reveal the presence of objects
with J-Ks color bluer than $\sim$0.8 mag, the color of a typical
irregular galaxy.
On the basis of the criteria we used to discriminate between
star
and galaxy (\S 3.2) those of them brighter than Ks$\sim$20 turned 
out to be stars.
The remaining blue sources  have  been classified  
as galaxies since they have either a FWHM larger than a point source or
a  stellarity index lower than 0.9.
However, almost all of them (14 in the CDF and 8 in the HDFS)
are fainter than Ks=21 where  star/galaxy classification is unfeasible  in our data.
Three out of the eight HDFS blue sources lie in the WFPC2 field: while two of them
are pointlike in the F814W band image, the remaining one displays an irregular shape.  

The first hypothesis we could make about the nature of these sources is that they are 
Galactic objects.
In this case, a possible explanation for the blue IR color of at 
least some of them
is that they  are extreme M sub-dwarfs, 
similar to those studied by Leggett et al. (1998).  
These sources have metallicities as low as 
about one-hundredth solar, are characterized by 
effective temperatures 
T$_{eff}\sim3000$ K and by masses ranging 0.09-0.15 M$_\odot$.
The objects analyzed by Leggett et al. have  magnitudes in the range 9$<K<15$,
 colors J-Ks$\ge0.4$ mag
 and absolute magnitudes in the range  9$<M_K<10$.
Most of our extremely blue sources are in fact in this color range.
Under the hypothesis that they are M sub-dwarfs with absolute magnitudes
similar to those of Leggett et al., 
the present observations should have reached  $\sim6$ kpc considering
a limiting magnitude  Ks$\sim 22.5$. 

Sources even bluer than J-Ks=0.4 and approaching J-K$\sim0$ can be consistent 
with  T-type brown dwarfs (e.g. Nakajima et al. 1995; Cuby et al. 1999).
These IR extremely blue  objects   are characterized by red  
optical-IR colors.
For instance, the methane brown dwarf NTTDF J1205-0744 (Cuby et al. 1999),
the most distant reported to date, 
has a magnitude Ks=20.3. and  colors J-K$\sim0$, I-K$>6$.
We reveal 3 faint sources in our samples with such a blue IR color: 2 in the CDF sample
having  J-Ks$\leq0.3$  and 1 in the HDFS sample with  J-Ks$\leq0.1$.
All of them are fainter than Ks=22 where the  
uncertainty in the color  estimate  is larger than 0.3 mag. 
If our three sources are methane brown dwarfs they have
to be 2.5 times more distant than NTTDF J1205-0744 (considering $\Delta K=2$ magnitudes).
The analysis of the optical data will help to assess the nature of these sources.

The second hypothesis about the nature of our extremely blue sources is that
they are extragalactic objects.
Using  Bruzual \& Charlot models we can account for colors less blue
than  0.65, the case of a 1 Gyr old irregular  galaxy at $z=0$ 
with a constant star formation (1 M$_\odot$/yr) and a metallicity lower than 
0.2Z$_\odot$.
Bluer colors can be obtained only assuming irregulars  younger than 1Gyr.   
Another possibility is that these blue objects are  
  $z>3$ sources  with strong UV excess, like those reported by Sullivan et al. (2000).
They debate the possibility that the UV light in these objects is
 due to a non-thermal
source such as QSO/AGN perhaps superimposed on a star-forming component.
In this case we would observe the rest-frame UV ($<3000$ \AA) in the J band 
and  wavelengths short-ward of $5500$ \AA~~ in the Ks band.
In fact, by red-shifting the mean energy distribution (MED) derived by Elvis et al. (1994)
for quasars we are not able to produce colors bluer than 0.85 ($z\sim3$).
The large dispersion about this MED results in 
an uncertainty we estimate to be about $0.2-0.3$ mag.
However, it is worth noting that  QSOs showing colors as blue as J-K$\sim0.6-0.5$
are observed also at redshift $z<3$ (Francis et al. 2000).   
}

\section{Summary and conclusions}
We presented counts and colors of galaxies detected in the deep
J and Ks images obtained with the near-IR camera ISAAC at the ESO
VLT telescope centered on the Chandra Deep Field and on the Hubble
Deep Field-South.
The co-addition of short dithered 2.5$\times$2.5 arcmin images
led to a total exposure time of about 8 hours in Ks  and 
to a limiting surface brightness Ks$\simeq$22.8 mag/arcsec$^2$
and J$\simeq$24.5 mag/arcsec$^2$ on both the fields.

We found an excellent agreement between the counts derived in the two
fields.
On the other hand our counts and their slope lie in between those
previously derived by other authors at comparable depth.
In particular we estimate a slope of $\gamma_J$=0.34 and 
$\gamma_K$=0.28 at J$>20$ and Ks$>19$ respectively.
We do not observe a turnoff in the counts down to the faintest magnitudes.

The observed J and Ks galaxy counts can be well fitted in a flat universe 
($q_0=0.5$ or $\lambda_0>0$)  only assuming  models 
including some degree of merging.
These models were previously found also to give a good fit  to the optical
number counts of the HDFS.
On the contrary models not including merging  are not able to 
reproduce the observed counts, independent of the cosmology.

Our estimated lower limits to the fraction of $z>1$ galaxies
based on the J-K$>1.9$ color selection criterion are
broadly consistent with the predictions for hierarchical models
as well as for pure luminosity evolution models.

We identify 7 sources in the CDF (or 2$\%$ of the sample) and 20 source 
in the HDFS (or 5$\%$ of the sample) with colors redder than J-Ks=2.3
and magnitudes $20<Ks<22$, equivalent to a surface density of 1.2 and
2.7 sources per square arcmin respectively.
Their extreme red colors suggest that they are old ellipticals 
or dusty star forming galaxies at redshift $z>2-3$.
However some of the most compact of them could be  very low-mass
stars which can display very red near-IR colors.
Another possibility, at least for the faintest and optically undetected ones, 
is that they are $z>10$ Lyman break galaxies.

A further analysis of their near-IR properties combined with those at
optical wavelength will allow us to constraint their nature.

\begin{acknowledgements}
We thank Emanuele Bertone for the useful checks
he made on the photometric properties of the different filters. 
We would like to acknowledge helpful discussions about blue sources
with Paola Severgnini, Marcella Longhetti and Davide Rizzo. 
We acknowledge the support of the network "Formation and Evolution of
Galaxies" set up by the European Commission under contract ERB
FMRX-CT96-086 of its TMR programme and of the ASI contract ARS-98-226.
The data here presented have been obtained as
part of an ESO Service Mode programme.

\end{acknowledgements}


\begin{thebibliography}{}
\bibitem{} Arnouts S., D'Odorico S., Cristiani S., Zaggia S., Fontana A.,
Giallongo E., 1999, A\&A 341, 641
\bibitem{} Babul A., Rees M., 1992, MNRAS 255, 346
\bibitem{} Balbi A., et al. 2000, astro-ph/0005124
\bibitem{} Baugh C. M., Cole S., Frank C. S., 1996, MNRAS 283, 1361
\bibitem{} Baugh, C. M.; Benson, A. J.; Cole, S.; Frenk, C. S.; Lacey, C. G.,
1999, MNRAS 305, L21
\bibitem{} Bernstein, G. M., Nichol, R. C., Tyson, J. A., Ulmer, M. P., 
Wittman, D. 1995, AJ, 110, 1507
\bibitem{} Bershady, M. A., Lowenthal, J. D., Koo D. C. 1998, ApJ 505, 50
\bibitem{} Bershady, M. A., Subbarao M., Koo D. C., Szalay A., Kron R. G.,
1999, in preparation
\bibitem{} Bertin, E., Arnouts, S., 1996, A\&AS, 117, 393 
\bibitem{} Broadhurst T. J., Ellis R. S., Glazebrook K., 1992, Nat 355, 55
\bibitem{} Bruzual A. G., Margis C. G., Calvert N., 1988, ApJ 333, 673
\bibitem{} Bruzual A. G., Charlot S., 1993, ApJ 405, 538
\bibitem{} Buzzoni A., 1989, ApJS 71, 817
\bibitem{} Buzzoni A., 1995, ApJS 98, 69
\bibitem{} Chabrier G., Baraffe I., Allard F., Haushildt P., 2000, ApJ 542, 464
\bibitem{} Cowie, L. L., Gardner, J. P., Lilly S. J., McLean I., 1990, ApJ 360, L1
\bibitem{} Cowie, L. L., Gardner, J. P., Hu, E. M., Songaila, A., Hodapp, K.
W., Wainscoat, R. J. 1994, ApJ, 434, 114
\bibitem{} Cuby J. G., Saracco P., Moorwood A. F. M., D'Odorico S., Lidman C., Comeron F.,
Spyromilio J., 1999, A\&A 349, L41
\bibitem{} Daddi E., Cimatti A., Pozzetti L., Hoekstra H., Röttgering H. J. A., Renzini A., Zamorani G., Mannucci F., 2000, A\&A 361, 535
\bibitem{} Dahn C. C., et al., 2000, in {\em From Giant Planets to Cool Stars},
ed. M. Marley and C. Griffith, in press
\bibitem{} De Bernardis P. et al. 2000, astro-ph/0011469
\bibitem{} De Propris, R., Eisenhardt, P. R., Stanford, S. A., Dickinson, M. 1998a, ApJ, 503
\bibitem{} Dickinson M., Hanley C., Elston R., et al., 2000, ApJ 531, 624
\bibitem{} Djorgovski,  S., et al. 1995, ApJ, 438, L13
\bibitem{} Eisenhardt P., Elston R., Stanford S. A., et al., 2000, (astro-ph/0002468)
\bibitem{} Fontana A., Menci N., D'Odorico S., et al., 1999, MNRAS 310, L27
\bibitem{} Fontana A., D'Odorico S., Poli F., et al., 2000, AJ 120, 2206
\bibitem{} Francis P. J., Whiting M. T., Webster R. L. 2000, PASA 17, 56
\bibitem{} Gardner, J. P., Cowie, L. L., Wainscoat, R. J. 1993, 
ApJ, 415, L9
\bibitem{} Gardner, J. P., Sharples, M. R.,  Carrasco, B. E., Frenk, C. S., 1997, MNRAS 282, L1
\bibitem{} Gardner, J. P., Sharples, M. R., Frenk, C. S., Carrasco, B. E. 1997,
ApJ, 480, L99
\bibitem{} Gardner, J. P., 1998, PASP 110, 291
\bibitem{} Glazebrook, K., Peacock, J., Collins, C., Miller, L. 1994, MNRAS, 266, 65
\bibitem{} Gronwall C., Koo D. C., 1995, ApJ 440, L1
\bibitem{} Huang J.-S., Cowie L. L., Gardner J. P., Hu E. M., Songaila A., Wainscoat R. J. 1997, ApJ 476, 12
\bibitem{} Kauffmann G., 1996, MNRAS 281, 487
\bibitem{} Kauffmann G., Charlot S., 1998, MNRAS 297, L23
\bibitem{} Le Fevre O., Abraham R., Lilly S. J., et al., 2000, MNRAS 311, 565
\bibitem{} Leggett S. K., Allard F., Hauschildt P. H., 1998, ApJ 509, 836
\bibitem{} Lobo, C., Biviano, A., Durret, F., Gerbal, D., Le Fevre, O., Mazure, A., Slezak, E. 1997, A\&A, 317, 385
\bibitem{} Maihara T., Iwamuro F., Tanabe H., et al., 2000, PASJ in press, (astro-ph/0009409)
\bibitem{} Mamon G. A., 1998, in {\em Wide field surveys in cosmology}, XIV IAP Meeting,
eds Colombi S., Mellier Y., Raban B., p. 323
\bibitem{} Martini P., 2000, AJ submitted, (astro-ph/0008328)
\bibitem{} Marzke, R. O., et al. 1994, AJ, 108, 436
\bibitem{} Marzke, R. O., Da Costa, L. N. 1997, AJ, 113, 1
\bibitem{} McCarthy P., Carlberg R., Marzke R., et al. 2000, (astro-ph/0011499)
\bibitem{} McCracken H. J., Metcalfe N., Shanks T., Campos A., Gardner J. P., Fong R., 2000, MNRAS 311, 707
\bibitem{} McLeod, B. A., Bernstein, G. M., Reike, M. J., Tollestrup, E. V., Fazio, G. G.,  1995, ApJS, 96, 117
\bibitem{} Metcalfe N.,  Shanks T., Campos A., McCracken H. J., Fong R., 2000, MNRAS in press (astro-ph/0010153)
\bibitem{} Minezaki, T., Kobayashi, Y., Yoshii, Y., Peterson, B. A., 1998,
ApJ, 494, 111
\bibitem{} Molinari E., Chincarini G., Moretti  A., De Grandi S., 1998, A\&A 338, 874
\bibitem{} Moorwood A. F. M. , Cuby J. G., Ballester P., et al.,  1999, The Messenger 95, 1
\bibitem{} Moustakas, L. A., Davis, M., Graham, J. R., Silk, J., Peterson, B. A., Yoshii, Y., 1997, ApJ, 475, 445
\bibitem{} Nakajima T., Oppenheimer B. R., Kulkarni S. R., Golimowski D. A., Matthews K., 
Durrance S. T. 1995, Nature 378, 463 
\bibitem{} Peacock J. A., 1987, in {em Astrophysical Jets and their Engines}, 
ed. Kundt W., 171, Reidel Dordrecht
\bibitem{} Perlmutter S., et al., 1999, ApJ 517, 565
\bibitem{} Persson E., Murphy D. C., Krzeminski W., Roth M., Rieke M. J., 1998, AJ 166, 2475
\bibitem{} Pozzetti L., Bruzual A. G., Zamorani G., 1996, MNRAS 281, 953
\bibitem{} Pozzetti L., Madau P., Zamorani G., Ferguson H., Bruzual A. G., 1998, MNRAS 298, 1133
\bibitem{} Riess A. G., et al., 1998, AJ 116, 1009
\bibitem{} Roche N., Eales S. A., Hippelein H., Willot C. J., 1999, MNRAS 306, 358
\bibitem{} Rocca-Volmerange B., Guiderdoni B., 1990, MNRAS 247, 166
\bibitem{} Saracco P., Iovino, A., Garilli, B., Maccagni, D., Chincarini, G.,
 1997, AJ, 114, 887
\bibitem{} Saracco P., D'Odorico S., Moorwood A. F. M., Buzzoni A., Cuby J. G., Lidman C., 1999, A\&A 349, 751
\bibitem{} Saracco P., et al., 2001, in preparation
\bibitem{} Sullivan M., Treyer M. A., Ellis R. S., Bridges T. J., Milliard B., 
Donas J., 2000, MNRAS 312, 442
\bibitem{} Szokoly, G. P., Subbarao, M. U., Connolly, A. J., Mobasher, B. 1998, ApJ, 492, 452
\bibitem{} Teplitz H. I., Gardner J. P., Malumuth E. M., Heap S. R., 1998, ApJ 507, L17
\bibitem{} Vaisanen P., Tollestrup E. V., Willner S. P., Cohen M., 2000, ApJ in press (astro-ph/0004224)
\bibitem{} Vanzella E., Cristiani S., Saracco P., Arnouts S., Fontana A., Giallongo E., Grazian A., Iovino A., 2001, AJ, in preparation
\bibitem{} Volonteri M., Saracco P., Chincarini G., Bolzonella M., A\&A 362, 487
\bibitem{} Wang B., 1991, ApJ 383, L37
\bibitem{} Yahata N., Lanzetta K. M., Chen H-W., et al. 2000, ApJ 538, 493
\bibitem{} Zucca, E., et al. 1997, A\&A, 326, 477
\end{thebibliography}
\end{document}